\documentclass[aps,prd,groupedaddress,showpacs,superscriptaddress,twocolumn,eqsecnum,floatfix]{revtex4-1}
\usepackage{graphicx}
\usepackage{dcolumn}
\usepackage{bm}
\begin{document}
\title{The decay $J/\psi\to\gamma\phi\phi$: spin dependence of amplitude and angular distributions of photons with linear polarizations.}
\author{A.~A.~Kozhevnikov}
\email[]{kozhev@math.nsc.ru} \affiliation{Laboratory of
Theoretical Physics, S.~L.~Sobolev Institute for Mathematics, Novosibirsk, Russian
Federation}
\affiliation{Novosibirsk State University, Novosibirsk, Russian
Federation}
\date{\today}
\begin{abstract}
Based on the effective invariant amplitudes of the transitions $J/\psi\to\gamma X(J^P)$ and $X(J^P)\to\phi\phi$, the spin dependence of the $J/\psi\to\gamma X(J^P)\to\gamma\phi\phi$ decay amplitude is obtained in the case of the intermediate resonances $X$ with $J^P=0^\pm,1^\pm,2^\pm$. Angular distributions of  photons   with  the definite linear  polarizations relative to the plane spanned by the momentum of initial electron and the total momentum of the $\phi\phi$ pair in the reaction $e^+e^-\to J/\psi\to\gamma X(J^P)\to\gamma\phi\phi$ are calculated. It is shown that the sign of the asymmetry of the  distributions of the photons polarized in the above plane and orthogonal to it correlates with the signature $P_X(-1)^{J_X}$ of the $X$ resonance with given spin-parity $J_X^{P_X}$ so it may help to establish this quantum number in a way which does not depend on the specific model of the $X(J^P)\to\phi\phi$ amplitude.
\end{abstract}
\maketitle
\section{Introduction}
\label{intro}
The study of decay  $J/\psi\to\gamma\phi\phi$ \cite{bisello1986,bai1990,ablikim2008,BES16} is of interest in view of  the possible  existence of exotic glueball states  decaying into the $\phi\phi$ pair \cite{Linden,Etkin85,Booth86,Etkin88}. The spin-parity quantum numbers of resonance states decaying into $\phi\phi$ are reported to be $J^P=0^+ , 0^-$, and $2^+$ \cite{ablikim2008,BES16,PDG}. However, the unit spin assignment cannot be excluded for all states with masses $m_X<m_{\eta_c}$ because the  $\gamma\gamma$ decay mode ruling out the state $J=1$ \cite{landau1948,yang50} is seen only in the case of  the $f_2(2300)$ resonance \cite{PDG,abe2004,uehara2013}. Note that  the spin-parity assignment $J^{PC}=1^{-+}$ is of special interest since it is forbidden in the $q\bar q$ model, hence the resonance with such quantum numbers  is explicitly exotic differently from the one with the quantum numbers $J^{PC}=0^{\pm +}$, $1^{++}$, and $2^{\pm +}$ which can be either the $q\bar q$ state or crypto-exotic one \cite{chang79}.

As it was pointed out, the $\phi\phi$ decay mode could be useful for the determination of the parity of resonances decaying to the above final state \cite{chang79,chang78,trueman78}. The basic idea was to exploit an analog of the Yang test \cite{yang50a} by measuring the  distribution over the angle between the decay planes spanned by the momenta of $K^+$ and $K^-$ mesons from the decay $\phi\to K^+K^-$. But, in the Yang test, the decay planes of $e^+e^-$ pairs from the decay chain $\pi^0\to\gamma\gamma$, $\gamma\to e^+e^-$,  are clearly distinguishable. In the meantime, the final $\phi$ mesons in the decay $X\to\phi\phi$ are relatively slow,  hence  there is strong interference between the amplitudes $M(X\to\phi\phi\to K^+_{p_1}K^-_{p_2}K^+_{p_3}K^-_{p_4})$ with the permutations of the kaon momenta $p_1\leftrightarrow p_3$ and $p_2\leftrightarrow p_4$ so that the $K^+K^-$ decay plane cannot be attributed to the specific $\phi$ meson.  A possible way to overcome this difficulty was suggested in Ref.~\cite{chang79}. It is based on the idea to determine the parity of the $\phi\phi$ state by making the direction-polarization correlation measurements with the linearly polarized photons  in the  reaction $e^+e^-\to J/\psi\to\gamma X\to\gamma\phi\phi$.

In the present work, we study theoretically a different way to exploit the linearly polarized photons for the purpose of the signature determination of  resonances in $\phi\phi$ system. To this end, based on the effective invariant amplitudes of the transitions $J/\psi\to\gamma X(J^P)$ and $X(J^P)\to\phi\phi$, the angular distribution is calculated of the final photons in the reaction $e^+e^-\to J/\psi\to\gamma\phi\phi$ which are linearly polarized relative to the plane spanned by the $e^+e^-$ beam direction and the total momentum of the $\phi\phi$ pair. The resulting expressions are given in terms  of independent helicity amplitudes $M_{\lambda_{J/\psi},\lambda_\gamma,\lambda_X}$  of the decay $J/\psi\to\gamma X(J^P)$ \cite{zou03} and in terms of the amplitudes of the decay $X(J^P)\to\phi\phi$ with given total spin and orbital momentum of the $\phi\phi$ state. The cases of the spin-parity and  charge conjugation assignments $J^{PC}=0^{\pm+},1^{\pm+},$ and $2^{\pm+}$ of the resonance $X$ are considered explicitly. Following Ref.~\cite{BES16}, the process
$J/\psi\to\gamma X(J^{PC})\to\gamma\phi\phi$   is considered here to be dominant.

Kinematic notations are the following. The 4-momentum assignment is $$J/\psi(Q)\to\gamma(k)X(q)\to\gamma(k)\phi(k_1)\phi(k_2);$$ $\epsilon_\mu$,  $\epsilon^{(X)}_\mu$,  $\epsilon_{1\mu}$, $\epsilon_{2\mu}$ (${\bm\xi}$, ${\bm\xi}^{(X)}$,  ${\bm\xi}_1$, ${\bm\xi}_2$) are,
respectively, the polarization four-vectors of the $J/\psi$ meson, intermediate spin one resonance $X$,   and $\phi$ mesons (their three-dimensional counterparts in their respective rest frames);  $e_\mu=(0,{\bm e})$ stands for the polarization four-vector of the photon, $\epsilon_{\mu\nu\lambda\sigma}$ is the Levy-Civita tensor and $e_{abc}$ is its three-dimensional counterpart. Corresponding polarization tensors of the spin two intermediate resonance $X$ are denoted as $T_{\mu\nu}$, $t_{ij}$.
The energy-momentum four-vector of the $\phi\phi$ state in the $J/\psi$ rest frame is $q=(q_0,{\bm q})$,
\begin{eqnarray}\label{qmu}
q_0&=&\frac{m^2_{J/\psi}+m^2_{12}}{2m_{J/\psi}},\nonumber\\
{\bm q}&=&-{\bm k}=-{\bm n}\frac{m^2_{J/\psi}-m^2_{12}}{2m_{J/\psi}},\end{eqnarray}where ${\bm n}$ stands for the unit vector in the  photon direction of motion, and $m_{12}$ is invariant mass of the $\phi\phi$ pair. The energy-momentum of one of $\phi$ mesons, $k_{1\mu}=(k^\ast_{10},{\bm k}^\ast_1)$, in the $\phi\phi$ center-of-mass system, is
\begin{eqnarray}\label{k1mu}
k^\ast_{10}&=&\frac{1}{2}m_{12},\nonumber\\
{\bm k}^\ast_1&=&\frac{{\bm n}_1}{2}\sqrt{m^2_{12}-4m^2_\phi},
\end{eqnarray}with ${\bm n}_1$ being the unit vector in the $\phi$ meson direction of motion. In what follows,
\begin{widetext}
\begin{eqnarray}\label{D}
D_{X(J^P)}(m^2_{12})&=&m^2_{X(J^P)}-m^2_{12}-im_{12}\left[\Gamma_{X(J^P)\to\phi\phi}(m_{12})+\sum\Gamma_{X(J^P)\to{\rm non}\phi\phi}(m_{12})\right]
\end{eqnarray}
\end{widetext}
will stand for the inverse propagator of the resonance $X(J^P)$; the energy dependence of the $\phi\phi$ decay mode is included explicitly while width $\Gamma_{X(J^P)\to{\rm non}\phi\phi}(m_{12})$ takes into account other possible decay modes.

Further material is organized as follows. The $J/\psi\to\gamma X(J^P)\to\gamma\phi\phi$ decay amplitudes for the $X$ resonance with spin-parities $J^P=0^\pm$, $1^\pm$, and $2^\pm$ are given in Sect.~\ref{sec1}. Angular distributions  of the linearly polarized photons in the reaction $e^+e^-\to J/\psi\to\gamma X(J^P)\to\gamma\phi\phi$ and their asymmetry are calculated in Sect.~\ref{sec2}. Section \ref{sec3} is devoted to the numerical estimates of the magnitude of above asymmetry using the results of fits \cite{kozh19} of the data  \cite{BES16} on the reaction $e^+e^-\to J/\psi\to\gamma\phi\phi$. Discussion and concluding remarks are given in Sect.~\ref{sec4}. Expressions for  invariant amplitudes with their three-dimensional counterparts, the partial widths of the decays $X(J^P)\to\phi\phi$,  the procedure of derivation of the averaging over $\phi\phi$ mass spectrum, and some additional details are given in appendices.

\section{Amplitudes}\label{sec1}
~

Here we follow the method of obtaining the expressions for amplitudes outlined in Ref.~\cite{kozh19}. Schematically, each amplitude, $M_{J/\psi\to\gamma X(J^P)}$ and $M_{X(J^P)\to\phi\phi}$ is written in the respective rest frame, $J/\psi$ and $X(J^P)$, so that the sum over polarizations (if any) of the intermediate resonance $X(J^P)$ can be fulfilled simply in terms of the Kronecker delta symbols
\begin{equation}\label{polsum1}
\sum_{\lambda_X}\xi_i^{(X)}\xi_j^{(X)}=\delta_{ij}
\end{equation}in case of $X(1^\pm)$ and
\begin{equation}\label{polsum2}
\sum_{\lambda_X}t_{ij}t_{kl}=\frac{1}{2}(\delta_{ik}\delta_{jl}+\delta_{il}\delta_{jk})-\frac{1}{3}\delta_{ij}\delta_{kl}\equiv\Pi_{ij,kl}
\end{equation}in case of $X(2^\pm)$. Let us give the expressions for the amplitudes $M^{(J^P)}\equiv M_{J/\psi\to\gamma X(J^P)\to\gamma\phi\phi}$ in terms of independent helicity amplitudes $M^{(J^P)}_{\lambda_{J/\psi},\lambda_\gamma,\lambda_X}$ of the decay $J/\psi\to\gamma X(J^P)$ \cite{zou03} and in terms of the $X(J^P)\to\phi\phi$ decay amplitudes $f^{(J^P)}_{SL}$ with given spin $S$ and orbital angular momentum $L$ of the $\phi\phi$ state. Necessary expressions for the invariant amplitudes $M_{J/\psi\to\gamma X(J^P)}$ and $M_{X(J^P)\to\phi\phi}$ with the lowest admissible powers of momenta and their respective three-dimensional counterparts side by side with the relations expressing $M^{(J^P)}_{\lambda_{J/\psi},\lambda_\gamma,\lambda_X}$ and $f^{(J^P)}_{SL}$  via effective coupling constants and kinematic factors are given in Appendix \ref{app0}. One has
\begin{eqnarray}\label{ampssc}
M^{(0^+)}&=&\frac{M^{(0^+)}_{1,1,0}({\bm\xi}{\bm e})}{D_{X(0^+)}(m^2_{12})}\left[f^{(0^+)}_{00}\delta_{ab}+f^{(0^+)}_{22}n_{1a}n_{1b}\right]\times\nonumber\\&&\xi_{1a}\xi_{1b}
\end{eqnarray} in the case of $J^P=0^+$;
\begin{eqnarray}\label{amps0mi}
M^{(0^-)}&=&\frac{iM^{(0^-)}_{1,1,0}f^{(0^-)}_{11}}{D_{X(0^-)}(m^2_{12})}({\bm\xi}[{\bm e}\times{\bm n}])e_{abc}n_{1c}\xi_{1a}\xi_{2b}
\end{eqnarray}in the case of $J^P=0^-$;
\begin{eqnarray}\label{ampsax}
M^{(1^+)}&=&\frac{f^{(1^+)}_{22}}{D_{X(1^+)}(m^2_{12})}\left\{M^{(1^+)}_{1,1,0}({\bm\xi}[{\bm e}\times{\bm n}])
n_c-\right.\nonumber\\&&\left.M^{(1^+)}_{0,1,1}({\bm\xi}{\bm n})[{\bm e}\times{\bm n}]_c\right\}(e_{cad}n_{1b}+e_{cbd}n_{1a})\times\nonumber\\&&n_{1d}\xi_{1a}\xi_{1b}
\end{eqnarray}in the case of $J^P=1^+$;
\begin{eqnarray}\label{ampsvec}
M^{(1^-)}&=&\frac{f^{(1^-)}_{11}}{D_{X(1^-)}(m^2_{12})}\left\{M^{(1^-)}_{1,1,0}({\bm\xi}{\bm e})
[{\bm n}\times{\bm n}_1]_c-\right.\nonumber\\&&\left.M^{(1^-)}_{0,1,1}({\bm\xi}{\bm n})[{\bm e}\times{\bm n}_1]_c\right\}
e_{abc}\xi_{1a}\xi_{1b}
\end{eqnarray}in the case of $J^P=1^-$;
\begin{eqnarray}\label{a2pl}
M^{(2^+)}&=&\left\{\frac{1}{2}\left(\sqrt{6}M^{(2^+)}_{1,1,0}-
M^{(2^+)}_{-1,1,2}\right)({\bm\xi}\cdot{\bm e})n_in_j-\right.\nonumber\\&&\left.
M^{(2^+)}_{-1,1,2}\xi_{\bot i}e_j-\sqrt{2}M^{(2^+)}_{0,1,1}({\bm\xi}\cdot{\bm n})e_in_j\right\}\times\nonumber\\&&\left[f^{(2^+)}_{20}\delta_{ka}\delta_{lb}+f^{(2^+)}_{02}
\delta_{ab}n_{1k}n_{1l}+\right.\nonumber\\&&\left.
f^{(2^+)}_{22}(n_{1a}\delta_{kb}+n_{1b}\delta_{ka})n_{1l}+\right.\nonumber\\&&\left.f^{(2^+)}_{24}n_{1a}n_{1b}n_{1k}n_{1l}\right]
\frac{\Pi_{ij,kl}\xi_{1a}\xi_{2b}}{D_{X(2^+)}(m^2_{12})}.
\end{eqnarray}in the case of $J^P=2^+$;
\begin{eqnarray}\label{a2mi}
M^{(2^-)}&=&i\left\{\sqrt{2}M^{(2^-)}_{0,1,1}({\bm\xi}{\bm n})[{\bm n}\times{\bm e}]_in_j+\sqrt{\frac{3}{2}}M^{(2^-)}_{1,1,0}\times\right.\nonumber\\&&\left.
({\bm n}[{\bm\xi}_\bot\times{\bm e}])n_in_j+
\frac{1}{2}M^{(2^-)}_{-1,1,2}\left(\xi_{\bot i}[{\bm n}\times{\bm e}]_j+\right.\right.\nonumber\\&&\left.\left.
[{\bm n}\times{\bm\xi}]_ie_j\right)\right\}\left(f^{(2^-)}_{11}e_{kab}
+f^{(2^-)}_{13}e_{abc}n_{1c}n_{1k}\right)\times\nonumber\\&&\frac{\Pi_{ij,kl}n_{1l}}{D_{X(2^+)}(m^2_{12})}\times\xi_{1a}\xi_{2b}.
\end{eqnarray}in the case of $J^P=2^-$. The quantity $D_{X(J^P)}(m^2_{12})$ in the $J/\psi\to\gamma X(J^P)\to\gamma\phi\phi$ decay amplitudes Eqs.~(\ref{ampssc}), (\ref{amps0mi}), (\ref{ampsax}), (\ref{ampsvec}), (\ref{a2pl}), and (\ref{a2mi}) is the inverse propagator of the $X(J^P)$ resonance given by Eq.~(\ref{D}).

Using  Eqs.~(\ref{ampssc}), (\ref{amps0mi}), (\ref{ampsax}),  (\ref{ampsvec}), (\ref{a2pl}), and (\ref{a2mi}) for the specific $X(J^P)$ resonance  contribution one can write the full amplitude as the sum of these expressions and study the coherence properties of these contributions to the total $J/\psi\to\gamma\phi\phi$ decay width. The direct calculation of the final probability distribution shows that, when summed over polarizations of the final $\phi$ mesons but keeping fixed their direction of motion, the vanishing ones are all interference terms  with opposite space parities.  The nonvanishing ones are $(0^+-2^+)$, $(0^--2^-)$, $(1^--2^-)$, and $(1^+-2^+)$. Among them, the $(0^+-2^+)$ and  $(0^--2^-)$ interference terms are proportional to $n_{1k}n_{1l}-\delta_{kl}/3$. They vanish after integration over $\phi\phi$ phase space because the averaging over the $\phi$ meson direction of motion is fulfilled with the help of the relation
\begin{equation}\label{nn}
\left\langle n_{1i}n_{1j}\right\rangle\equiv\int n_{1i}n_{1j}\frac{d\Omega_{{\bm n}_1}}{4\pi}=\frac{1}{3}\delta_{ij}.
\end{equation}
The $(1^+-2^+)$ interference term, after summation over polarizations of both $\phi$ mesons, is proportional to $({\bm e}\cdot{\bm n}_1)n_kn_{1l}-({\bm n}\cdot{\bm n}_1)e_kn_{1l}$ which, after integration over ${\bm n}_1$ with the use of Eq.~(\ref{nn}) reduces to $e_kn_l-e_nn_k$ and also vanishes after multiplication by the tensor $\Pi_{ij,kl}$  which is symmetric in both pairs of indices,  $ij$ and $kl$. The $(1^--2^-)$ interference term contains two structures, $[{\bm n}\times{\bm n}_1]_kn_{1l}$ and  $[{\bm e}\times{\bm n}_1]_kn_{1l}$. After  integration over ${\bm n}_1$  with the help of Eq.~(\ref{nn}) they reduce to, respectively, $\epsilon_{akl}n_a$ and $\epsilon_{akl}e_a$  and vanish being multiplied by the tensor $\Pi_{ij,kl}$ symmetric in $kl$. So, after summation over polarizations and integration over ${\bm n}_1$, the probability distribution 
$$\frac{d\sigma_{e^+e^-\to J/\psi\to\gamma\phi\phi}}{d\Omega_{{\bm n}}dm_{12}}$$
can be represented as the incoherent sum of contributions with the different quantum numbers.

\section{Angular distributions for linearly polarized photons}\label{sec2}
~

Let us find the  angular distribution of the polarized photons in the reaction $e^+e^-\to J/\psi\to\gamma X\to\gamma\phi\phi$. To be specific, we will use for the above quantity the expression for the differential decay width of the transversally polarized $J/\psi$ meson assuming the single $X$ resonance in the intermediate state. The direct calculation (see Appendix \ref{app2} for the nontrivial particular case) shows that in all cases of our interest it can be written in the form of the averaging over $\phi\phi$ mass spectrum,
\begin{widetext}
\begin{eqnarray}\label{phgen}
\frac{d\Gamma_{J/\psi\to\gamma X\to\gamma\phi\phi}}{d\Omega_\gamma}&=&\frac{1}{\pi}\int_{4m^2_\phi}^{m^2_{J/\psi}}
\frac{m_{12}\Gamma_{X\to\phi\phi}(m^2_{12})}{|D_X(m^2_{12})|^2}\times
\frac{d\Gamma_{J/\psi\to\gamma X}}{d\Omega_\gamma}dm^2_{12}
\equiv\left\langle\frac{d\Gamma_{J/\psi\to\gamma X}}{d\Omega_\gamma}\right\rangle,
\end{eqnarray}
\end{widetext}
where it is assumed that both $\Gamma_{J/\psi\to\gamma X\to\gamma\phi\phi}$ and $\Gamma_{J/\psi\to\gamma X}$ correspond to the partial widths with the given polarization states of the $J/\psi$ meson and photon (that is, neither the averaging over the $J/\psi$ meson spin projection nor the summation over the photon polarization are fulfilled). In what follows, the shorthand notation
\begin{eqnarray}\label{short}
a^{(J^P)}_{\lambda_{J/\psi},\lambda_\gamma,\lambda_X}&\equiv&\frac{1}{\pi}\int_{4m^2_\phi}^{m^2_{J/\psi}}
\frac{m_{12}\Gamma_{X(J^P)\to\phi\phi}(m^2_{12})}{|D_{X(J^P)}(m^2_{12})|^2}\times\nonumber\\&&
\left|M^{(J^P)}_{\lambda_{J/\psi},\lambda_\gamma,\lambda_X}\right|^2|{\bm k}|dm^2_{12}
\end{eqnarray}
will be used where $M^{(J^P)}_{\lambda_{J/\psi},\lambda_\gamma,\lambda_X}$ are the helicity amplitudes of the radiative decay $J/\psi\to\gamma X(J^P)$ to the $X$ resonance with given spin-parity $J^P$ and with the specific helicities of the particles involved. See Eqs.~(\ref{notsc}), (\ref{notps}), (\ref{notax}), (\ref{notvec}),  (\ref{hel2pl}), and (\ref{hel2mi}).

Let us specify the coordinate system. The electrons move along the $z$ axis, the plane spanned by the momenta $({\bm p}_{e^-},{\bm k})$ is the $xz$ plane, so that ${\bm n}=(\sin\theta_\gamma,0,\cos\theta_\gamma)$. Of our primer interest are the angular distributions over the polar angle of the photon $\theta_\gamma$, in case of its linear polarizations in the reaction $e^+e^-\to J/\psi\to\gamma\phi\phi$. The vectors of two linear polarization states in  this system are chosen in the form
\begin{eqnarray}\label{e12}
{\bm e}_1&=&(0,1,0),\nonumber\\
{\bm e}_2&=&(-\cos\theta_\gamma,0,\sin\theta_\gamma).
\end{eqnarray}
They have the properties ${\bm e}^2_{1,2}=1$,
\begin{equation}\label{nee}
[{\bm e}_1\times{\bm e}_2]={\bm n},
\end{equation} and other two relations obtained by the cyclic permutations of the latter. As is evident from the above expressions, ${\bm e}_1$ is orthogonal to the $({\bm p}_{e^-},{\bm k})$ plane while ${\bm e}_2$ lies in this plane. One can equivalently characterize the above plane as the plane spanned by the momentum of initial electron and the total momentum of the $\phi\phi$ pair.

Since the $J/\psi$ meson is produced in $e^+e^-$ annihilation, it has only two spin projections $\lambda_{J/\psi}=\pm1$ on the direction of motion of the initial electron. This will be taken into account using the summation rule
\begin{equation}\label{sumpsi}
\sum_{\lambda_{J/\psi}=\pm1}\xi_i^{(\lambda_{J/\psi})}\xi_j^{(\lambda_{J/\psi})}=\delta_{ij}-\delta_{i3}\delta_{j3}.
\end{equation}
We calculate the asymmetry in the form
\begin{widetext}
\begin{eqnarray}\label{asym}
A^{(J^P)}(\theta_\gamma)&=&\left(\left\langle\frac{d\Gamma^{(1)}_{J/\psi\to\gamma X(J^P)}}{d\Omega_\gamma}\right\rangle-
\left\langle\frac{d\Gamma^{(2)}_{J/\psi\to\gamma X(J^P)}}{d\Omega_\gamma}\right\rangle\right)
\left(\left\langle\frac{d\Gamma^{(1)}_{J/\psi\to\gamma X(J^P)}}{d\Omega_\gamma}\right\rangle+
\left\langle\frac{d\Gamma^{(2)}_{J/\psi\to\gamma X(J^P)}}{d\Omega_\gamma}\right\rangle\right)^{-1},
\end{eqnarray}
\end{widetext}
where the superscript (1,2) refers to the photon polarization vector ${\bm e}_{1,2}$, respectively. Notice that the quantity in the denominator of Eq.~(\ref{asym}) is proportional to the probability to emit unpolarized photon.

Let us go to the presentation of the angular distributions of the linearly polarized photons. In what follows, the shorthand notation
\begin{equation}\label{notang}
\frac{d\Gamma^{(J^P)}}{d\Omega_\gamma}\equiv\frac{d\Gamma_{J/\psi\to\gamma X(J^P)\to\gamma\phi\phi}}{d\Omega_\gamma}.
\end{equation}
is used. It is also assumed that the sum over $\lambda_{J/\psi}$ goes over $\pm1$.

(a) $X=0^+$.  One obtains from Eq.~(\ref{ampssc}) that
\begin{equation}\label{dis0pl1}
\sum_{\lambda_{J/\psi}}|M_{J/\psi\to\gamma X(0^+)}|^2=\left|M^{(0^+)}_{1,1,0}\right|^2({\bm e}^2-e^2_z).
\end{equation}The angular distribution of the linearly polarized photons is written in the form
\begin{eqnarray}\label{dis0pl2}
\frac{d\Gamma^{(0^+)}}{d\Omega_\gamma}&=&Ba^{(0^+)}_{1,1,0}\times\left\{\begin{array}{c}
 1\mbox{, }{\bm e}={\bm e}_1,\\
 \cos^2\theta_\gamma\mbox{, }{\bm e}={\bm e}_2.\\
        \end{array}
\right.
\end{eqnarray}
Hereafter
\begin{equation}\label{norm}
B=\frac{1}{(8\pi m_{J/\psi})^2}
\end{equation}is the normalization factor which  includes the division by four due to both the averaging over two transverse polarizations of the $J/\psi$ meson produced in $e^+e^-$ annihilation and the identity of the final $\phi$ mesons.
The asymmetry Eq.~(\ref{asym}) is
\begin{equation}\label{asym0pl}
A^{(0^+)}(\theta_\gamma)=\frac{\sin^2\theta_\gamma}{1+\cos^2\theta_\gamma}>0.
\end{equation}

(b) $X=0^-$. The sum over  polarizations and the angular distribution in the present case look, respectively, like
\begin{eqnarray}\label{dis0mi1}
\sum_{\lambda_{J/\psi}}|M_{J/\psi\to\gamma X(0^-)}|^2&=&\left|M^{(0^-)}_{1,1,0}\right|^2([{\bm n}\times{\bm e}]^2-\nonumber\\&&
[{\bm n}\times{\bm e}]^2_z)
\end{eqnarray}and
\begin{eqnarray}\label{dis0ml2}
\frac{d\Gamma^{(0^-)}}{d\Omega_\gamma}&=&Ba^{(0^-)}_{1,1,0}\times\left\{\begin{array}{c}
\cos^2\theta_\gamma\mbox{, }{\bm e}={\bm e}_1,\\
 1\mbox{, }{\bm e}={\bm e}_2.\\
        \end{array}
\right.
\end{eqnarray} The asymmetry Eq.~(\ref{asym}) in the present case is
\begin{equation}\label{asym0mi}
A^{(0^-)}(\theta_\gamma)=-\frac{\sin^2\theta_\gamma}{1+\cos^2\theta_\gamma}<0.
\end{equation}The signs of the asymmetry of the linearly polarized photons are opposite for the scalar and pseudoscalar $X$ resonances.  The normalized angular distribution of unpolarized photons is
\begin{equation}\label{npol0}
\frac{dW_{0^\pm}}{d\cos\theta_\gamma}=\frac{3}{8}(1+\cos^2\theta_\gamma).
\end{equation}It is the same for the spinless $X$ resonances with opposite space parities.

(c) $X=1^+$. In this case, the sum over  spin projections  is
\begin{eqnarray}\label{dis1pl1}
\sum_{\lambda_X,\lambda_{J/\psi}}|M_{J/\psi\to\gamma X(1^+)}|^2&=&\left|M^{(1^+)}_{1,1,0}\right|^2([{\bm n}\times{\bm e}]^2-\nonumber\\&&
[{\bm n}\times{\bm e}]^2_z)+\left|M^{(1^+)}_{0,1,1}\right|^2\times\nonumber\\&&
(1-n^2_z)[{\bm n}\times{\bm e}]^2.
\end{eqnarray}The angular distribution of the linearly polarized photons and the asymmetry Eq.~(\ref{asym}) look, respectively,  as follows:
\begin{widetext}
\begin{eqnarray}\label{dis1pl2}
\frac{d\Gamma^{(1^+)}}{d\Omega_\gamma}&=&B\times\left\{\begin{array}{c}
a^{(1^+)}_{1,1,0}\cos^2\theta_\gamma+a^{(1^+)}_{0,1,1}\sin^2\theta_\gamma\mbox{, }{\bm e}={\bm e}_1\\
a^{(1^+)}_{1,1,0}+a^{(1^+)}_{0,1,1}\sin^2\theta_\gamma \mbox{, }{\bm e}={\bm e}_2\\
        \end{array}
\right.,\nonumber\\
A^{(1^+)}(\theta_\gamma)&=&-a^{(1^+)}_{1,1,0}\sin^2\theta_\gamma\times\left[a^{(1^+)}_{1,1,0}(1+\cos^2\theta_\gamma)+
2a^{(1^+)}_{0,1,1}\sin^2\theta_\gamma\right]^{-1}<0.
\end{eqnarray}
\end{widetext}

(d) $X=1^-$. The corresponding sum over  spin projections  is
\begin{eqnarray}\label{dis1mi1}
\sum_{\lambda_X,\lambda_{J/\psi}}|M_{J/\psi\to\gamma X(1^-)}|^2&=&\left|M^{(1^-)}_{1,1,0}\right|^2({\bm e}^2-e^2_z)+\nonumber\\&&
\left|M^{(1^-)}_{0,1,1}\right|^2(1-n^2_z)\times\nonumber\\&&{\bm e}^2.
\end{eqnarray}
The angular distribution of the linearly polarized photons and the asymmetry Eq.~(\ref{asym}) look, respectively,  as follows:
\begin{widetext}
\begin{eqnarray}\label{dis1mi2}
\frac{d\Gamma^{(1^-)}}{d\Omega_\gamma}&=&
B\times\left\{\begin{array}{c}
a^{(1^-)}_{1,1,0}+a^{(1^-)}_{0,1,1}\sin^2\theta_\gamma \mbox{, }{\bm e}={\bm e}_1,\\
a^{(1^-)}_{1,1,0}\cos^2\theta_\gamma+a^{(1^-)}_{0,1,1}\sin^2\theta_\gamma\mbox{, }{\bm e}={\bm e}_2,\\
        \end{array}
\right.\nonumber\\
A^{(1^-)}(\theta_\gamma)&=&a^{(1^-)}_{1,1,0}\sin^2\theta_\gamma\left[a^{(1^-)}_{1,1,0}(1+\cos^2\theta_\gamma)+
2a^{(1^-)}_{0,1,1}\sin^2\theta_\gamma\right]^{-1}>0.
\end{eqnarray}
\end{widetext}
The signs of the asymmetry are opposite for the vector and axial vector $X$ resonances while the angular distribution  of unpolarized photons,
\begin{equation}\label{npol1}
\frac{dW_{1^\pm}}{d\cos\theta_\gamma}=\frac{3[a^{(1^\pm)}_{1,1,0}(1+\cos^2\theta_\gamma)+2a^{(1^\pm)}_{0,1,1}\sin^2\theta_\gamma]}
{8[a^{(1^\pm)}_{1,1,0}+a^{(1^\pm)}_{0,1,1}]},
\end{equation}is the same in cases of the spin one $X$ resonances with opposite space parities. One can observe that in the cases of spin zero and spin one resonances  the expressions for $\sum_{\lambda_X,\lambda_{J/\psi}=\pm1}|M_{J/\psi\to\gamma X}|^2$ for opposite space parities are related by the interchange ${\bm e}\leftrightarrow[{\bm n}\times{\bm e}]$. This is not the case for the tensor resonances. See below.

(e) $X=2^+$. The modulus squared of the $J/\psi\to\gamma X(2^+)$ transition amplitude  summed over polarizations of the tensor $X$ resonance is
\begin{eqnarray}\label{sum2pl}
\sum_{\lambda_X}\left|M_{J/\psi\to\gamma X(2^+)}\right|^2&=&\left|M_{1,1,0}^{(2^+)}\right|^2({\bm\xi}\cdot{\bm e})^2+\nonumber\\&&\left|M_{0,1,1}^{(2^+)}\right|^2
({\bm\xi}\cdot{\bm n})^2{\bm e}^2+\nonumber\\&&
\frac{1}{2}\left|M_{-1,1,2}^{(2^+)}\right|^2{\bm\xi}_\bot^2{\bm e}^2,
\end{eqnarray}
so that
\begin{eqnarray}\label{sum2pla}
\sum_{\lambda_X,\lambda_{J/\psi}}\left|M_{J/\psi\to\gamma (2^+)}\right|^2&=&\left|M_{1,1,0}^{(2^+)}\right|^2({\bm e}^2-e^2_z)+\nonumber\\&&\left|M_{0,1,1}^{(2^+)}\right|^2(1-n^2_z)^2{\bm e}^2+\nonumber\\&&
\frac{1}{2}\left|M_{-1,1,2}^{(2^+)}\right|^2(1+n^2_z)\times\nonumber\\&&{\bm e}^2.
\end{eqnarray}
Using this expression and Eq.~(\ref{e12}) one obtains the angular distributions and asymmetry:
\begin{widetext}
\begin{eqnarray}\label{dis2pl2}
\frac{d\Gamma^{(2^+)}}{d\Omega_\gamma}&=&B\left[a^{(2^+)}_{0,1,1}\sin^2\theta_\gamma+
\frac{1}{2}a^{(2^+)}_{-1,1,2}(1+\cos^2\theta_\gamma)+
a^{(2^+)}_{1,1,0}\times
\left\{\begin{array}{c}
1\mbox{, }{\bm e}={\bm e}_1,\\
\cos^2\theta_\gamma\mbox{, }{\bm e}={\bm e}_1,\\
        \end{array}
\right.\right],\nonumber\\
A^{(2^+)}(\theta_\gamma)&=&a^{(2^+)}_{1,1,0}\sin^2\theta_\gamma\left[\left(a^{(2^+)}_{1,1,0}+a^{(2^+)}_{-1,1,2}
\right)(1+\cos^2\theta_\gamma)+2a^{(2^+)}_{0,1,1}\sin^2\theta_\gamma\right]^{-1}>0.
\end{eqnarray}
\end{widetext}

(f) $X=2^-$. The modulus squared of the $J/\psi\to\gamma X(2^-)$ decay amplitude  summed over polarizations of the tensor $X$ resonance and this quantity summed over the transverse polarization states of the $J/\psi$ meson are, respectively,
\begin{eqnarray}\label{sum2mi}
\sum_{\lambda_X}\left|M_{J/\psi\to\gamma X(2^-)}\right|^2&=&\left|M_{1,1,0}^{(2^-)}\right|^2({\bm\xi}\cdot[{\bm e}\times{\bm n}])^2+\nonumber\\&&\left|M_{0,1,1}^{(2^-)}\right|^2({\bm\xi}\cdot{\bm n})^2[{\bm e}\times{\bm n}]^2+\nonumber\\&&
\frac{1}{4}\left|M_{-1,1,2}^{(2^-)}\right|^2\left(\frac{1}{2}{\bm\xi}_\bot^2[{\bm e}\times{\bm n}]^2+\right.\nonumber\\&&\left.\frac{1}{2}{\bm e}^2[{\bm n}\times{\bm\xi}]^2+({\bm\xi}\cdot[{\bm e}\times{\bm n}])^2+\right.\nonumber\\&&\left.({\bm\xi}\cdot{\bm e})^2\right),
\end{eqnarray}and
\begin{eqnarray}\label{sum2mia}
\sum_{\lambda_X,\lambda_{J/\psi}}\left|M_{J/\psi\to\gamma X(2^-)}\right|^2&=&\left|M^{(2^-)}_{1,1,0}\right|^2\left([{\bm e}\times{\bm n}]^2-\right.\nonumber\\&&\left.
[{\bm e}\times{\bm n}]^2_z\right)+\left|M_{0,1,1}^{(2^-)}\right|^2\times\nonumber\\&&
(1-n^2_z)[{\bm e}\times{\bm n}]^2+\nonumber\\&&
\frac{1}{4}\left|M^{(2^-)}_{-1,1,2}\right|^2\left[\frac{1}{2}(3+n^2_z)\times\right.\nonumber\\&&\left.
\left([{\bm e}\times{\bm n}]^2+{\bm e}^2\right)-\right.\nonumber\\&&\left.
[{\bm e}\times{\bm n}]^2_z-e^2_z\right].
\end{eqnarray}
Note that  differently from the spin zero and spin one cases, there are the terms in Eqs.~(\ref{sum2pl}) and (\ref{sum2mi}) that are not related by the interchange ${\bm e}\leftrightarrow[{\bm n}\times{\bm e}]$. One finds from  Eq.~(\ref{sum2mia}) the angular distributions of the linearly polarized photons with ${\bm e}={\bm e}_1$ and ${\bm e}={\bm e}_2$ and their asymmetry the following expressions:
\begin{widetext}
\begin{eqnarray}\label{dis2mi2}
\frac{d\Gamma_1^{(2^-)}}{d\Omega_\gamma}&=&B\left[a^{(2^-)}_{0,1,1}\sin^2\theta_\gamma+
\left(a^{(2^-)}_{1,1,0}+\frac{1}{4}a^{(2^-)}_{-1,1,2}\right)
\cos^2\theta_\gamma+\frac{1}{4}a^{(2^-)}_{-1,1,2}
(2+\cos^2\theta_\gamma)\right],\nonumber\\
\frac{d\Gamma_2^{(2^-)}}{d\Omega_\gamma}&=&B\left[a^{(2^-)}_{0,1,1}\sin^2\theta_\gamma+a^{(2^-)}_{1,1,0}+
\frac{1}{4}a^{(2^-)}_{-1,1,2}+
\frac{1}{4}a^{(2^-)}_{-1,1,2}(2+\cos^2\theta_\gamma-\sin^2\theta_\gamma)\right],\nonumber\\
A^{(2^-)}(\theta_\gamma)&=&-a^{(2^-)}_{1,1,0}\sin^2\theta_\gamma\left[\left(a^{(2^-)}_{1,1,0}+
a^{(2^-)}_{-1,1,2}\right)
(1+\cos^2\theta_\gamma)+2a^{(2^-)}_{0,1,1}\sin^2\theta_\gamma\right]^{-1}<0.
\end{eqnarray}
\end{widetext}
The normalized angular distributions of the unpolarized photons in the present case is
\begin{widetext}
\begin{eqnarray}\label{npol2}
\frac{dW_{2^\pm}}{d\cos\theta_\gamma}&=&\frac{3}{8}\times\left[(a^{(2^\pm)}_{1,1,0}+a^{(2^\pm)}_{-1,1,2})(1+\cos^2\theta_\gamma)
+2a^{(2^\pm)}_{0,1,1}\sin^2\theta_\gamma\right]\times
\left[a^{(2^\pm)}_{1,1,0}+a^{(2^\pm)}_{-1,1,2}+a^{(2^\pm)}_{0,1,1}\right]^{-1}.
\end{eqnarray}
\end{widetext}
It looks the same in cases of the spin two $X$ resonances with opposite space parities.

To summarize this section, one can say that, with the chosen orientations of the linear polarizations of the photon (orthogonal to the plane spanned by direction of the $e^+e^-$ beam axes and the momentum ${\bm k}$ or lying in this plane), the sign of the asymmetry Eq.~(\ref{asym}) coincides with the signature $\sigma_X=P_X(-1)^{J_X}$ of the $X$ resonance with spin-parity $J_X^{P_X}$.

\section{Asymmetry estimates using  available data}\label{sec3}
~

\begin{figure}
\resizebox{0.55\textwidth}{!}{%
  \includegraphics{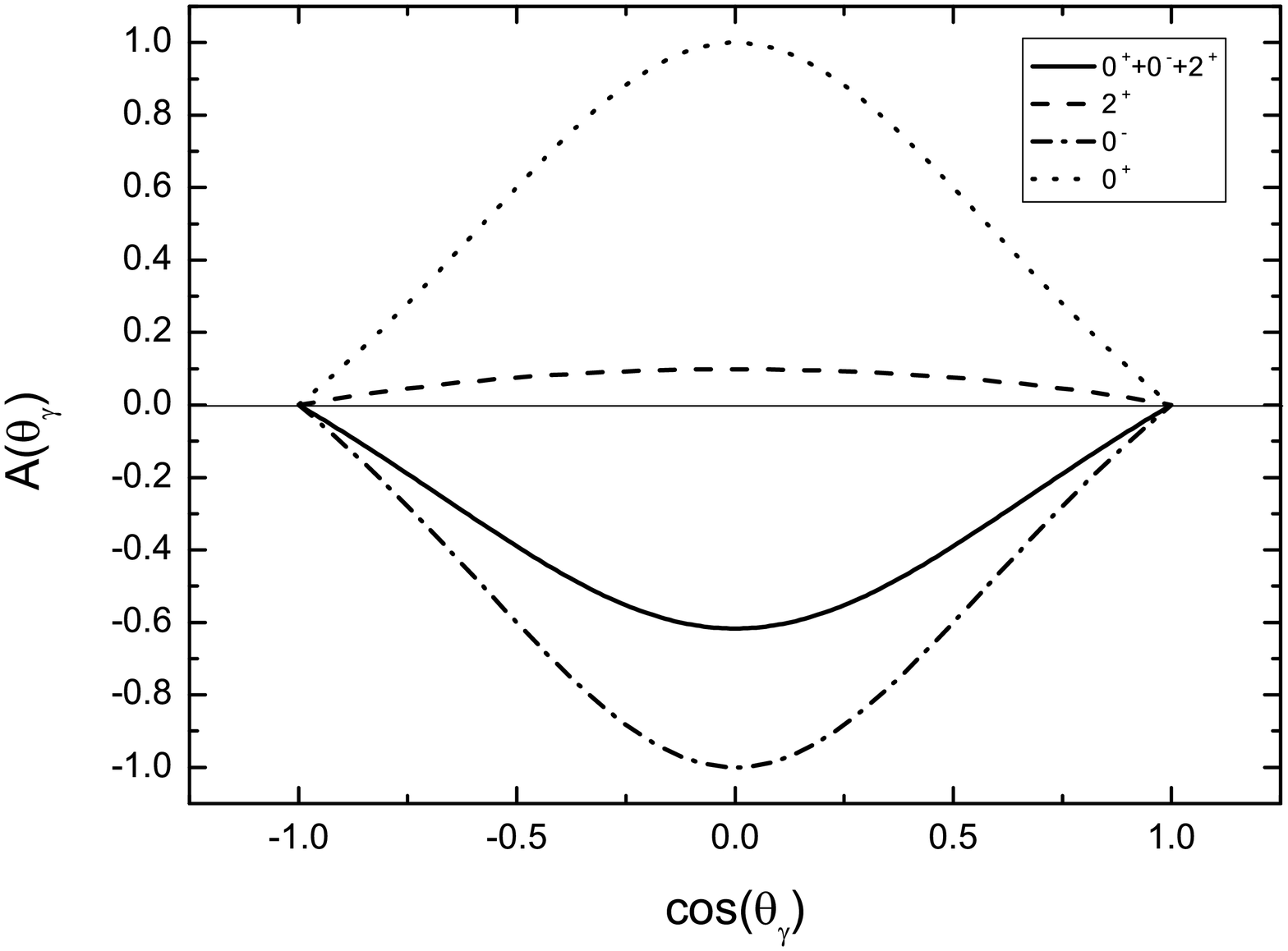}
}
\caption{Asymmetry (\ref{asymmix}) in the angular distributions of the linearly polarized photons in the reaction $e^+e^-\to J/\psi\to\gamma\phi\phi$ calculated using the resonance parameters extracted \cite{kozh19} from the BESIII data \cite{BES16}.}
\label{asymth}       
\end{figure}

Let us estimate the possible magnitude of asymmetry using the recent  data \cite{BES16} on disentangling the  $X(J^P)$ resonance contributions to the $e^+e^-\to J/\psi\to\gamma X(J^P)\to\gamma\phi\phi$ reaction amplitude. The authors of Ref.~\cite{BES16} restricted their analysis by the resonances with quantum numbers $0^\pm$ and $2^+$. As is evident from Eqs.~(\ref{asym0pl}) and (\ref{asym0mi}), in case of the spin zero resonances the asymmetry shown with the dotted and dash-dotted lines in Fig.~\ref{asymth} does not depend on the dynamical details. Further treatment requires the knowledge of the  magnitudes of coupling constants parameterizing invariant amplitudes. They can be found in Ref.~\cite{kozh19}. Taking into account the fact that the $0^+$ and $0^-$ contributions do not interfere, see the last paragraph in Sec.~\ref{sec1}, the asymmetry due to the combined spin zero component is
\begin{eqnarray}\label{asy0}
A^{(0^++0^-)}(\theta_\gamma)&=&\frac{\sin^2\theta_\gamma}{1+\cos^2\theta_\gamma}\times\frac{N^{0^+}-N^{0^-}}{N^{0^+}+N^{0^-}}=\nonumber\\&&
-\frac{0.84\sin^2\theta_\gamma}{1+\cos^2\theta_\gamma}.\end{eqnarray}
Hereafter $N^{(0^+)}=63$ and $N^{(0^-)}=708$ are the central values of the relative production rates of the scalar and pseudoscalar components of the $\phi\phi$ mass spectrum found in Ref.~\cite{kozh19} when fitting the data Ref.~\cite{BES16}.

The asymmetry $A^{(2^+)}(\theta_\gamma)$,  in case of the pure $2^+$ resonance  evaluated with the parameters obtained in Ref.~\cite{kozh19} is shown in Fig.~\ref{asymth} with the dashed line. Its maximum value is
\begin{equation}\label{asymmax}
A^{(2^+)}_{\rm max}=\frac{a^{(2^+)}_{1,1,0}}{a^{(2^+)}_{1,1,0}+a^{(2^+)}_{-1,1,2}+2a^{(2^+)}_{0,1,1}}=0.10\pm0.02.
\end{equation}It is useful to evaluate the integral asymmetry
$A^{(2^+)}_{\rm int}=(\Gamma_1-\Gamma_2)/(\Gamma_1+\Gamma_2)$ where $\Gamma_{1,2}$ corresponding to the two linear polarizations ${\bm e}_{1,2}$ are obtained from Eq.~(\ref{dis2pl2}) by integrating over solid angle. The result is
\begin{eqnarray}\label{asymint}
A^{(2^+)}_{\rm int}&=&\frac{1}{2}\times\frac{a^{(2^+)}_{1,1,0}}{a^{(2^+)}_{1,1,0}+a^{(2^+)}_{-1,1,2}+a^{(2^+)}_{0,1,1}}=\nonumber\\&&
0.07\pm0.01,\end{eqnarray}
to be compared with $A^{(0^\pm)}_{\rm int}=\pm1/2$ evaluated in case of the $0^\pm$ resonances.

The above estimates were obtained in case of pure contributions with given spin. One can go further and evaluate the asymmetry in case when the contribution of the sum of the $X(0^-)$, $X(0^+)$, $X(2^+)$ resonances is taken into account. As is shown in Sec.~\ref{sec1}, the interference between them disappears after integration over $\phi\phi$ mass spectrum so that the asymmetry in this case looks like
\begin{eqnarray}\label{asymmix}
A(\theta_\gamma)&=&\sin^2\theta_\gamma\left(N^{(0^+)}-N^{(0^-)}+\frac{{\cal N}a^{(2^+)}_{1,1,0}}{12\pi m^2_{J/\psi}}\right)\times\nonumber\\&&
\left\{(1+\cos^2\theta_\gamma)\left[N^{(0^+)}+N^{(0^-)}+\frac{{\cal N}}{12\pi m^2_{J/\psi}}\times\right.\right.\nonumber\\&&\left.\left.
\left(a^{(2^+)}_{1,1,0}+a^{(2^+)}_{-1,1,2}\right)\right]+\frac{{\cal N}a^{(2^+)}_{0,1,1}}{6\pi m^2_{J/\psi}}
\sin^2\theta_\gamma\right\}^{-1}.
\end{eqnarray}
The factor ${\cal N}$ in the above expression originates from the fact that only relative production rates $\Gamma_{J/\psi\to\gamma X(J^P)\to\gamma\phi\phi}$ found in Ref.~\cite{kozh19} can be extracted from the fits of the BESIII data \cite{BES16}. The dimension of ${\cal N}$ is GeV$^{-1}$; the products  ${\cal N}^{1/2}g^{(1,2,3)}_{J/\psi\gamma X(2^+)}$ for the radiative coupling constants necessary for numerical evaluation of the quantities $a^{(2^+)}_{\lambda_{J/\psi},\lambda_\gamma,\lambda_X}$ can be found in Ref.~\cite{kozh19}. The plots of  $A(\theta_\gamma)$  for some particular cases are shown in Fig.~\ref{asymth}.
%

\section{Discussion and conclusion}
\label{sec4}
~

The question of determining parity of the $X$  resonances with masses below $m_{\eta_c}$   produced in the decay chain $J/\psi\to\gamma X\to\gamma\phi\phi$ is crucial for establishing their nature \cite{chang79,chang78,trueman78}. The measurements of angular distributions of the circularly polarized photons and of the unpolarized ones have the same form for the resonances with opposite space  parities. However, as is shown in Sec.~\ref{sec2}, the angular distributions of the photons polarized in the plane spanned by the momenta of the electron and photon (equivalently, the total momentum of the $\phi\phi$ pair) in the reaction $e^+e^-\to J/\psi\to\gamma X\to\gamma\phi\phi$ and perpendicular to this plane are different. The signs of the asymmetry Eq.~(\ref{asym}) are model independent and are correlated with the signature $\sigma_X=P_X(-1)^{J_X}$ of the resonance $X$, providing the above decay chain is the dominant mechanism of production of the $\gamma\phi\phi$ state. With the choice corresponding to  Eqs.~(\ref{e12}) and (\ref{asym}), they are positive (negative) for the $X$ resonance with the signature $\sigma_X=+1(-1)$. This method of determination of the signature based on the analysis of angular distributions of the linearly polarized photons could be complement to the standard partial wave analysis since it does not depend on the specific model of the amplitude. However, the calculation  of the asymmetry magnitude requires the knowledge of the dynamical information. Note in this respect that it is not excluded that the coupling constants entering the effective invariant amplitudes $M_{X(J^P)\to\phi\phi}$ in Appendix \ref{app0} could be the functions of momenta describing sophisticated contributions, for example,  those described by the triangle diagrams \cite{liang19}. To be sure,  the transition $X(J^P)\to\phi\phi$ followed by the reaction $\phi\phi\to\phi\phi$ going via the $\eta$, $\eta^\prime$ exchanges, by the unitarity relation, results in appearance of imaginary parts of the triangle diagrams, hence the phases, of the effective couplings. However, the corresponding triangle diagrams do not possess the decay kinematics so the so called triangle singularities are absent in the process under consideration.

The method of measurement of the linear polarization of  photons by the $e^+e^-$ pair production in the field of heavy nuclei was proposed in Refs.~\cite{berlin50,wick51}. It is based on studying the correlation between the photon polarization plane and the $e^+e^-$ production plane \cite{max59}. The detailed study of this method was presented in Ref.~\cite{max62} where the calculation was done of the electron emission rate in its   dependence on the angle between the above planes. It was shown that under the specified kinematic conditions the preferred emission of   electrons is in the photon polarization plane \cite{max62}. In the context of the present work, it seems to be not impossible to detect the $e^+e^-$ pairs which are externally produced by the photons from the decay $J/\psi\to\gamma\phi\phi$. They can be detected with the tracking detection system to determine the momenta of electron and positron, hence the $e^+e^-$ production plane. Their total momentum should be equal to the total momentum of the four registered $K^+K^-K^+K^-$ mesons from the decay of the $\phi\phi$ pair.  The preferred emission of electrons in the direction orthogonal (parallel) to the plane spanned by the vectors $({\bm p}_{e^-},{\bm k})$ would point to the linear polarization ${\bm e}_1$ (${\bm e}_2$), respectively.

Of course, there is nothing special in the treatment of the particular case of the decay $J/\psi\to\gamma\phi\phi$ considered here. The undertaken task was motivated by available data \cite{BES16} on the decay $J/\psi\to\gamma\phi\phi$. The obtained results valid under assumption that the process proceeds predominantly as the two-step process $V\to\gamma X(J^P)$, $X(J^P)\to V_1V_1$, $J^P=0^\pm,1^\pm,2^\pm$, can be applied to any decay of
the type $V\to\gamma V_1V_1$, where $V$  are the vector mesons from the $\psi$ or $\Upsilon$ families, and $V_1=\rho,\omega,\phi$, with the evident replacements.

I am indebted to G.~Carboni for calling my attention to the possibility of the unit spin resonance in the $\phi\phi$ system. The work was supported by the program of fundamental scientific researches of the Siberian Branch of Russian Academy of Sciences no. II.15.1., project no. 0314-2019-0021.

\appendix

\section{The $J/\psi\to\gamma X(J^P)$ and $X(J^P)\to\phi\phi$ transition amplitudes.}\label{app0}
~

Let us give the expressions for the effective invariant amplitudes of the transitions $J/\psi\to\gamma X(J^P)$ and $X(J^P)\to\phi\phi$ with the lowest possible powers of momenta and their 3D counterparts valid in the gauge $e_0=0$, ${\bm e}{\bm n}=0$.

(a) $J^P=0^+$. One has

\begin{eqnarray}\label{ampssca}
M_{J/\psi\to\gamma X(0^+)}&=&-g^{(1)}_{J/\psi\gamma X(0^+)}(\epsilon e)=M^{(0^+)}_{1,1,0}({\bm\xi}{\bm e}),\nonumber\\
M_{X(0^+)\to\phi\phi}&=&-g^{(1)}_{X(0^+)\phi\phi}(\epsilon_1\epsilon_2)-\nonumber\\&&
g^{(2)}_{X(0^+)\phi\phi}(\epsilon_1k_2)(\epsilon_2k_1)=\nonumber\\&&
f^{(0^+)}_{00}({\bm\xi}_1{\bm\xi}_2)+\nonumber\\&&f^{(0^+)}_{22}({\bm\xi}_1{\bm n}_1)({\bm\xi}_2{\bm n}_1),
\end{eqnarray}
where
\begin{eqnarray}\label{notsc}
M^{(0^+)}_{1,1,0}&=&g^{(1)}_{J/\psi\gamma X(0^+)},\nonumber\\
f^{(0^+)}_{00}&=&g^{(1)}_{X(0^+)\phi\phi},\nonumber\\
f^{(0^+)}_{22}&=&\left[2g^{(1)}_{X(0^+)\phi\phi}+g^{(2)}_{X(0^+)\phi\phi}m^2_{12}\right]\frac{{\bm k}^{\ast2}_1}{m^2_\phi}.
\end{eqnarray}

(b) $J^P=0^-$. One has
\begin{eqnarray}\label{amps0mia}
M_{J/\psi\to\gamma X(0^-)}&=&g_{J/\psi\gamma X(0^-)}\epsilon_{\mu\nu\lambda\sigma}Q_\mu\epsilon_\nu k_\lambda e_\sigma=\nonumber\\&&
iM^{(0^-)}_{1,1,0}({\bm n}\cdot[{\bm\xi}\times{\bm e}]),\nonumber\\
M_{X(0^-)\to\phi\phi}&=&g_{X(0^-)\to\phi\phi}\epsilon_{\mu\nu\lambda\sigma}k_{1\mu}\epsilon_{1\nu}k_{2\lambda}\epsilon_{2\sigma}=\nonumber\\&&
g_{X(0^-)\phi\phi}m_{12}|{\bm k}^\ast_1|\times\nonumber\\&&({\bm n}_1\cdot[{\bm\xi}_1\times{\bm\xi}_2]),
\end{eqnarray}
where
\begin{eqnarray}
M^{(0^-)}_{1,1,0}&=&-ig_{J/\psi\gamma X(0^-)}m_{J/\psi}|{\bm k}|,\nonumber\\
f^{(0^-)}_{11}&=&g_{X(0^-)\to\phi\phi}m_{12}|{\bm k}^\ast_1|.
\label{notps}
\end{eqnarray}

(c) $J^P=1^+$. One has
\begin{eqnarray}\label{ampsaxa}
M_{J/\psi\to\gamma X(1^+)}&=&\epsilon_{\mu\nu\lambda\sigma}k_\mu e_\nu \left[g^{(1)}_{J/\psi\gamma X(1^+)}\epsilon_\lambda\epsilon^{(X)}_\sigma+
\right.\nonumber\\&&\left.g^{(2)}_{J/\psi\gamma X(1^+)}\epsilon_\lambda Q_\sigma\left(\epsilon^{(X)}Q\right)+\right.\nonumber\\&&\left.
g^{(3)}_{J/\psi\gamma X(1^+)}\epsilon^{(X)}_\lambda Q_\sigma(\epsilon k)\right]=\nonumber\\&&
iM^{(1^+)}_{1,1,0}({\bm\xi}[{\bm e}\times{\bm n}])\left({\bm\xi}^{(X)}{\bm n}\right)-\nonumber\\&&
iM^{(1^+)}_{0,1,1}\left({\bm\xi}^{(X)}[{\bm e}\times{\bm n}]\right)\left({\bm\xi}{\bm n}\right),\nonumber\\
M_{X(1^+)\to\phi\phi}&=&\epsilon_{\mu\nu\lambda\sigma}\epsilon^{(X)}_\mu\left\{g^{(1)}_{X(1^+)\phi\phi}
\epsilon_{1\nu}\epsilon_{2\lambda}\times\right.\nonumber\\&&\left.(k_1-k_2)_\sigma+g^{(2)}_{X(1^+)\phi\phi}
\left[\epsilon_{1\nu}(\epsilon_2k_1)-\right.\right.\nonumber\\&&\left.\left.
\epsilon_{2\nu}(\epsilon_1k_2)\right]k_{1\lambda}k_{2\sigma}\right\}=\nonumber\\&&
f^{(1^+)}_{22}\left\{\left({\bm\xi}_1\left[{\bm n}_1\times{\bm\xi}^{(X)}\right]\right)({\bm\xi}_2{\bm n}_1)+\right.\nonumber\\&&\left.
\left({\bm\xi}_2\left[{\bm n}_1\times{\bm\xi}^{(X)}\right]\right)({\bm\xi}_1{\bm n}_1)\right\},
\end{eqnarray}
where
\begin{eqnarray}\label{notax}
M^{(1^+)}_{1,1,0}&=&i\left[g^{(1)}_{J/\psi\gamma X(1^+)}k_0+\frac{m_{J/\psi}{\bm k}^2}{m_{12}}\times\right.\nonumber\\&&\left.
\left(\frac{2g^{(1)}_{J/\psi\gamma X(1^+)}}{m_{J/\psi}+m_{12}}+m_{J/\psi}g^{(2)}_{J/\psi\gamma X(1^+)}\right)\right],\nonumber\\
M^{(1^+)}_{0,1,1}&=&i\left(-g^{(1)}_{J/\psi\gamma X(1^+)}k_0+g^{(3)}_{J/\psi\gamma X(1^+)}{\bm k}^2\right),\nonumber\\
f^{(1^+)}_{22}&=&\left(2g^{(1)}_{X(1^+)\phi\phi}+m^2_{12}g^{(2)}_{X(1^+)\phi\phi}\right)\frac{{\bm k}_1^{\ast2}}{m_\phi},
\end{eqnarray}and $k_0=|{\bm k}|$.

(d) $J^P=1^-$. One has
\begin{eqnarray}\label{ampsveca}
M_{J/\psi\to\gamma X(1^-)}&=&g^{(1)}_{J/\psi\gamma X(1^-)}(\epsilon k)\left(e\epsilon^{(X)}\right)+\nonumber\\&&
g^{(2)}_{J/\psi\gamma X(1^-)}(\epsilon e)\left(\epsilon^{(X)}k\right)=\nonumber\\&&
-M^{(1^-)}_{0,1,1}({\bm\xi}{\bm n})({\bm\xi}^{(X)}{\bm e})+\nonumber\\&&M^{(1^-)}_{1,1,0}({\bm\xi}{\bm e})({\bm\xi}^{(X)}{\bm n}),\nonumber\\
M_{X(1^-)\to\phi\phi}&=&g_{X(1^-)\phi\phi}\left[(\epsilon_1k_2)(\epsilon_2\epsilon^{(X)})+\right.\nonumber\\&&\left.
(\epsilon_2k_1)(\epsilon_1\epsilon^{(X)})\right]=\nonumber\\&&
f^{(1^-)}_{11}\left([{\bm\xi}^{(X)}\times{\bm n}_1][{\bm\xi}_1\times{\bm\xi}_2]\right),
\end{eqnarray}
where
\begin{eqnarray}\label{notvec}
M^{(1^-)}_{0,1,1}&=&-|{\bm k}|g^{(1)}_{J/\psi\gamma X(1^-)},\nonumber\\
M^{(1^-)}_{1,1,0}&=&\frac{m_{J/\psi}|{\bm k}|}{m_{12}}g^{(2)}_{J/\psi\gamma X(1^-)},\nonumber\\
f^{(1^-)}_{11}&=&g_{X(1^-)\phi\phi}\frac{m_{12}|{\bm k}^{\ast}_1|}{m_\phi}.
\end{eqnarray}

(e) $J^P=2^+$. The amplitude $J/\psi\to\gamma X(2^+)$ looks like
\begin{eqnarray}\label{ampg2pl}
M_{J/\psi\to\gamma X(2^+)}&=&-\left[g^{(1)}_{J/\psi\gamma X(2^+)}(\epsilon e)Q_\mu Q_\nu+\right.\nonumber\\&&\left.
g^{(2)}_{J/\psi\gamma X(2^+)}(\epsilon k)e_\mu k_\nu+\right.\nonumber\\&&\left.
g^{(3)}_{J/\psi\gamma X(2^+)}\epsilon_\mu e_\nu\right]T_{\mu\nu}=\nonumber\\&&
\left[g_{02}({\bm\xi}\cdot{\bm e})n_in_j+g_{12}({\bm\xi}\cdot{\bm n})e_in_j+\right.\nonumber\\&&\left.
g_{20}\xi_ie_j\right]t_{ij},
\end{eqnarray}
where
\begin{eqnarray}\label{gab}
g_{02}&=&\frac{m^2_{J/\psi}{\bm k}^2}{m^2_{12}}g^{(1)}_{J/\psi\gamma X(2^+)},\nonumber\\
g_{12}&=&\frac{{\bm k}^2}{m_{12}}\left[g^{(2)}_{J/\psi\gamma X(2^+)}q_0+\frac{g^{(3)}_{J/\psi\gamma X(2^+)}}{q_0+m_{12}}\right],\nonumber\\
g_{20}&=&g^{(3)}_{J/\psi\gamma X(2^+)}.
\end{eqnarray}
The $X(2^+)\to\phi\phi$ amplitude can be written in the form
\begin{eqnarray}\label{tens2fi}
M_{X(2^+)\to\phi\phi}&=&\left\{g^{(1)}_{X(2^+)\phi\phi}\epsilon_{1\mu}\epsilon_{2\nu}+k_{1\mu}k_{2\nu}
\left[g^{(2)}_{X(2^+)\phi\phi}\times\right.\right.\nonumber\\&&\left.\left.
(\epsilon_1\epsilon_2)+g^{(3)}_{X(2^+)\phi\phi}(\epsilon_1k_2)(\epsilon_2k_1)\right]+\right.\nonumber\\&&\left.
g^{(4)}_{X(2^+)\phi\phi}\left[\epsilon_{1\mu}k_{2\nu}(\epsilon_2k_1)+\epsilon_{2\mu}k_{1\nu}(\epsilon_1k_2)\right]\right\}=\nonumber\\&&
\left\{f^{(2^+)}_{20}\xi_{1i}\xi_{2j}+f^{(2^+)}_{02}({\bm\xi}_1\cdot{\bm\xi}_2)n_{1i}n_{1j}+\right.\nonumber\\&&\left.
f^{(2^+)}_{22}\left[({\bm\xi}_1\cdot{\bm n}_1)\xi_{2i}+({\bm\xi}_2\cdot{\bm n}_1)\xi_{1i}\right]n_{1j}+\right.\nonumber\\&&\left.f^{(2^+)}_{24}({\bm\xi}_1\cdot{\bm n}_1)({\bm\xi}_2\cdot{\bm n}_1)n_{1i}n_{1j}\right\}t_{ij},
\end{eqnarray}
where
\begin{eqnarray}\label{fLS2pl}
f^{(2^+)}_{20}&=&g^{(1)}_{X(2^+)\phi\phi},\nonumber\\
f^{(2^+)}_{02}&=&g^{(2)}_{X(2^+)\phi\phi}{\bm k}_1^{\ast2},\nonumber\\
f^{(2^+)}_{22}&=&\frac{{\bm k}_1^{\ast2}}{m_\phi}\left[\frac{g^{(1)}_{X(2^+)\phi\phi}}{k^\ast_{10}+m_\phi}+m_{12}g^{(4)}_{X(2^+)\phi\phi}\right],\nonumber\\
f^{(2^+)}_{24}&=&\frac{{\bm k}_1^{\ast4}}{m^2_\phi}\left[\frac{g^{(1)}_{X(2^+)\phi\phi}}{(k^\ast_{10}+m_\phi)^2}+2g^{(2)}_{X(2^+)\phi\phi}+
\right.\nonumber\\&&\left.m^2_{12}g^{(3)}_{X(2^+)\phi\phi}+
\frac{2m_{12}g^{(4)}_{X(2^+)\phi\phi}}{k^\ast_{10}+m_\phi}\right].
\end{eqnarray}
Helicity amplitudes are expressed via the quantities Eq.~(\ref{gab}) as follows:
\begin{eqnarray}\label{hel2pl}
M^{(2^+)}_{1,1,0}&=&\frac{1}{\sqrt{6}}(2g_{02}-g_{20}),\nonumber\\
M^{(2^+)}_{0,1,1}&=&-\frac{1}{\sqrt{2}}(g_{12}+g_{20}),\nonumber\\
M^{(2^+)}_{-1,1,2}&=&-g_{20}.
\end{eqnarray}

(f) $J^P=2^-$. The vertex $J/\psi\to\gamma X(2^-)$ looks like
\begin{widetext}
\begin{eqnarray}\label{ampg2mi}
M_{J/\psi\to\gamma X(2^-)}&=&\epsilon_{\mu\nu\lambda\sigma}\left[g^{(1)}_{J/\psi\gamma X(2^-)}\epsilon_\mu e_\nu T_{\lambda\alpha}Q_\alpha k_\sigma+
g^{(2)}_{J/\psi\gamma X(2^-)}\epsilon_\mu e_\nu k_\lambda Q_\sigma T_{\alpha\beta}Q_\alpha Q_\beta+\right.\nonumber\\&&\left.
g^{(3)}_{J/\psi\gamma X(2^-)}e_\mu k_\nu T_{\lambda\alpha}\epsilon_\alpha Q_\sigma+
g^{(4)}_{J/\psi\gamma X(2^-)}(\epsilon k)e_\mu k_\nu T_{\lambda\alpha}Q_\alpha Q_\sigma\right]=\nonumber\\&&i\left\{\sqrt{2}M^{(2^-)}_{0,1,1}({\bm\xi}{\bm n})[{\bm n}\times{\bm e}]_in_j+
\sqrt{\frac{3}{2}}M^{(2^-)}_{1,1,0}({\bm n}[{\bm\xi}_\bot\times{\bm e}])n_in_j+\right.\nonumber\\&&\left.
\frac{1}{2}M^{(2^-)}_{-1,1,2}\left(\xi_{\bot i}[{\bm n}\times{\bm e}]_j+
[{\bm n}\times{\bm\xi}]_ie_j\right)\right\}t_{ij},
\end{eqnarray}
\end{widetext}
where ${\bm\xi}_\bot={\bm\xi}-{\bm n}({\bm\xi}{\bm n})$, and
\begin{eqnarray}\label{hel2mi}
M^{(2^-)}_{0,1,1}&=&\frac{im_{J/\psi}|{\bm k}|}{m_{12}\sqrt{2}}\left[-g^{(1)}_{J/\psi\gamma X(2^-)}k_0+
g^{(3)}_{J/\psi\gamma X(2^-)}q_0-\right.\nonumber\\&&\left.
m_{12}{\bm k}^2g^{(4)}_{J/\psi\gamma X(2^-)}\right],\nonumber\\
M^{(2^-)}_{1,1,0}&=&-i\frac{\sqrt{2}m_{J/\psi}|{\bm k}|}{\sqrt{3}m_{12}}\left[g^{(1)}_{J/\psi\gamma X(2^-)}k_0+\frac{m_{J/\psi}}{m_{12}}
\times\right.\nonumber\\&&\left.\left(\frac{2g^{(1)}_{J/\psi\gamma X(2^-)}}{m_{J/\psi}+m_{12}}-m_{J/\psi}g^{(2)}_{J/\psi\gamma X(2^-)}\right){\bm k}^2-
\right.\nonumber\\&&\left.\frac{m_{12}}{2}g^{(3)}_{J/\psi\gamma X(2^-)}\right],\nonumber\\
M^{(2^-)}_{-1,1,2}&=&im_{J/\psi}|{\bm k}|g^{(3)}_{J/\psi\gamma X(2^-)}.
\end{eqnarray}
The $X(2^-)\to\phi\phi$ vertex can be represented in the form
\begin{eqnarray}\label{tensmi2fi}
M_{X(2^-)\to\phi\phi}&=&\epsilon_{\mu\nu\lambda\sigma}\left\{g^{(1)}_{X(2^-)\phi\phi}q_\mu\epsilon_{1\nu}\epsilon_{2\lambda}
T_{\sigma\alpha}(k_1-k_2)_\alpha+\right.\nonumber\\&&\left.
g^{(2)}_{X(2^-)\phi\phi}k_{1\mu}\epsilon_{1\nu}k_{2\lambda}\epsilon_{2\sigma}T_{\alpha\beta}k_{1\alpha}k_{2\beta}+
\right.\nonumber\\&&\left.g^{(3)}_{X(2^-)\phi\phi}\left[(\epsilon_1k_2)\epsilon_{2\nu}k_{1\mu}k_{2\lambda}-\right.\right.\nonumber\\&&\left.\left.
(\epsilon_2k_1)\epsilon_{1\nu}k_{2\mu}k_{1\lambda}\right]T_{\sigma\alpha}(k_1-k_2)_\alpha+\right.\nonumber\\&&\left.
g^{(4)}_{X(2^-)\phi\phi}q_\mu\left(\epsilon_{1\nu}\epsilon_{2\alpha}k_{1\sigma}+\right.\right.\nonumber\\&&\left.\left.
\epsilon_{2\nu}\epsilon_{1\alpha}k_{2\sigma}\right)T_{\lambda\alpha}\right\}=\nonumber\\&&\left\{f^{(2^-)}_{11}[{\bm\xi}_1\times{\bm\xi}_2]_in_{1j}+
\right.\nonumber\\&&\left.f^{(2^-)}_{13}([{\bm\xi}_1\times{\bm\xi}_2]{\bm n}_1)
n_{1i}n_{1j}\right\}t_{ij},
\end{eqnarray}
where
\begin{eqnarray}\label{fLS2mi}
f^{(2^-)}_{11}&=&|{\bm k}^\ast_1|\left[g^{(2)}_{X(2^-)\phi\phi}-2g^{(1)}_{X(2^-)\phi\phi}\right],\nonumber\\
f^{(2^-)}_{13}&=&\frac{|{\bm k}^\ast_1|^3}{m_\phi}\left[2m_{12}g^{(3)}_{X(2^-)\phi\phi}-m_\phi g^{(2)}_{X(2^-)\phi\phi}+
\right.\nonumber\\&&\left.\frac{2g^{(1)}_{X(2^-)\phi\phi}-g^{(4)}_{X(2^-)\phi\phi}}{k^\ast_{10}+m_\phi}\right].
\end{eqnarray}

\section{The $X(J^P)\to\phi\phi$ decay widths}\label{app1}
~

Using the expressions for the $X(J^P)\to\phi\phi$ decay amplitudes it is straightforward to obtain the expressions for the energy dependence of the $\phi\phi$ decay widths for different spin-parities of the $X$ resonance. They are
\begin{eqnarray}\label{widths}
\Gamma_{X(0^-)\to\phi\phi}(m_{12})&=&\frac{g^2_{X(0^-)\to\phi\phi}}{8\pi}|{\bm k}^\ast_1|^3,\nonumber\\
\Gamma_{X(0^+)\to\phi\phi}(m_{12})&=&\frac{|{\bm k}^\ast_1|}{16\pi m^2_{12}}\left(2\left|f^{(0^+)}_{00}\right|^2+\right.\nonumber\\&&\left.
\left|f^{(0^+)}_{00}+f^{(0^+)}_{22}\right|^2\right),\nonumber\\
\Gamma_{X(1^-)\to\phi\phi}(m_{12})&=&\frac{g^2_{X(1^-)\to\phi\phi}}{12\pi m^2_\phi}|{\bm k}^\ast_1|^3,\nonumber\\
\Gamma_{X(1^+)\to\phi\phi}(m_{12})&=&\frac{|{\bm k}^\ast_1|}{12\pi m^2_{12}}\left|f^{(1^+)}_{22}\right|^2,\nonumber\\
\Gamma_{X(2^+)\to\phi\phi}(m_{12})&=&\frac{|{\bm k}^\ast_1|}{240\pi m^2_{12}}\left(10\left|f^{(2^+)}_{20}+f^{(2^+)}_{22}\right|^2+
\right.\nonumber\\&&\left.3\left|f^{(2^+)}_{20}\right|^2+2\left|f^{(2^+)}_{20}+f^{(2^+)}_{24}\right|^2+\right.\nonumber\\
&&\left.2\left|f^{(2^+)}_{02}+f^{(2^+)}_{24}\right|^2+\right.\nonumber\\&&\left.
4\left|f^{(2^+)}_{02}+f^{(2^+)}_{22}\right|^2+\right.\nonumber\\&&\left.
4\left|f^{(2^+)}_{22}+f^{(2^+)}_{24}\right|^2-4\left|f^{(2^+)}_{22}\right|^2-\right.\nonumber\\&&\left.
6\left|f^{(2^+)}_{24}\right|^2\right),\nonumber\\
\Gamma_{X(2^-)\to\phi\phi}(m_{12})&=&\frac{|{\bm k}^\ast_1|}{60\pi m^2_{12}}\left[\frac{3}{2}\left|f^{(2^-)}_{11}\right|^2+\right.\nonumber\\&&\left.
\left|f^{(2^-)}_{11}+f^{(2^-)}_{13}\right|^2\right].
\end{eqnarray}
The expressions for the amplitudes $f^{(0^+)}_{SL}$, $f^{(1^+)}_{SL}$ and $f^{(2^\pm)}_{SL}$ with definite spin $S$ and orbital angular momentum $L$ of the $\phi\phi$ state are given by Eqs.~(\ref{notsc}), (\ref{notax}), (\ref{fLS2pl}), and (\ref{fLS2mi}).

\section{Integrating over $\phi\phi$ phase space}\label{app2}
~

Let us deduce the expression Eq.~(\ref{phgen}) for the angular distribution of the final photon in the decay $J/\psi\to\gamma X\to\gamma\phi\phi$. This equation is evidently valid in case of the resonance $X$ with spin zero or one. The case of $J^P=2^+$ is considered in Ref.~\cite{kozh19}.  Let us consider here the particular case of the intermediate resonance $X=2^-$. We start with the expression \cite{PDG}
\begin{eqnarray}\label{dGam}
d\Gamma&=&\frac{1}{(2\pi)^516m^2_{J/\psi}}|M|^2|{\bm k}^\ast_1||{\bm k}|dm_{12}d\Omega^\ast_1d\Omega_\gamma,
\end{eqnarray}where $|M|^2=\sum_{\lambda_1,\lambda_2}|M_{J/\psi\to\gamma X(2^-)\to\gamma\phi\phi}|^2$,
for the differential partial width of the $J/\psi\to\gamma X(2^-)\to\gamma\phi\phi$ decay with the fixed polarizations of the $J/\psi$ meson and photon.  It is convenient to write  Eq.~(\ref{ampg2mi})
in the form $M_{J/\psi\to\gamma X(2^-)}=M_{ij}t_{ij}$. Then one obtains that
\begin{eqnarray}\label{appp1}
|M|^2&=&\frac{2M_{ij}M^\ast_{i^\prime j^\prime}\Pi_{ij,kl}\Pi_{i^\prime j^\prime,k^\prime l^\prime}}{|D_{X(2^-)}|^2}
\left[\left|f^{(2^-)}_{11}\right|^2\delta_{kk^\prime}n_{1l}n_{1l^\prime}+\right.\nonumber\\&&\left.
\left(\left|f^{(2^-)}_{11}+f^{(2^-)}_{13}\right|^2-\left|f^{(2^-)}_{11}\right|^2\right)\times\right.\nonumber\\&&\left.
n_{1k}n_{1l}n_{1k^\prime}n_{1l^\prime}\right].
\end{eqnarray}
The next step is to integrate over the solid angle $d\Omega^\ast_1$ of the $\phi$ meson with the help of the known expressions for the averaging over the solid angle. They are given by Eq.~(\ref{nn}) and the following expression:
\begin{eqnarray*}
\left\langle n_{1i}n_{1j}n_{1k}n_{1l}\right\rangle&=&\frac{1}{15}(\delta_{ik}\delta_{jl}+\delta_{il}\delta_{jk}+\delta_{ij}\delta_{kl}).
\end{eqnarray*}
Then, using the fact that $\Pi$ is the projector, $\Pi_{ij,kl}\Pi_{kl,mn}=\Pi_{ij,mn}$, one obtains that
\begin{eqnarray}\label{appp2}
\frac{d\Gamma_{J/\psi\to\gamma X(2^-)\to\gamma\phi\phi}}{d\Omega_\gamma dm_{12}}&=&\frac{M_{ij}\Pi_{ij,i^\prime j^\prime}M^\ast_{i^\prime j^\prime}|{\bm k}^\ast_1||{\bm k}|}{(2\pi)^460m^2_{J/\psi}|D_{X(2^-)}|^2}\times\nonumber\\&&
\left(\frac{3}{2}\left|f_{11}^{(2^-)}\right|^2+\right.\nonumber\\&&\left.
\left|f_{11}^{(2^-)}+f_{13}^{(2^-)}\right|^2\right).
\end{eqnarray}
Since the photon angular distribution in the $J/\psi\to\gamma X(2^-)$ decay  is
\begin{eqnarray*}
\frac{d\Gamma_{J/\psi\to\gamma X(2^-)}}{d\Omega_\gamma}&=&\frac{M_{ij}\Pi_{ij,i^\prime j^\prime}M^\ast_{i^\prime j^\prime}|{\bm k}|}{32\pi^2m^2_{J/\psi}}
\end{eqnarray*} and the $\Gamma_{X(2^-)\to\phi\phi}$ decay width is given by the last line in Eq.~(\ref{widths}),
the substitution of these expressions to Eq.~(\ref{appp2}) followed by integration over $m_{12}$ results in Eq.~({\ref{phgen}).

%
%

%
%
\end{document}